\documentclass[longauth]{aa}

\usepackage{graphicx}
\usepackage{xcolor}
\usepackage[switch]{lineno}
\usepackage{subfigure}
\usepackage{txfonts}
\begin{document} 

   \titlerunning{Molecular Cloud Paradigm Near SNR G106.3+2.7} 
   \title{Testing the molecular cloud paradigm for ultra-high-energy gamma ray emission from the direction of SNR G106.3+2.7}

   \author{ \small R.~Alfaro \inst{\ref{IF-UNAM}} \and C.~Alvarez \inst{\ref{UNACH}} \and J.C.~Arteaga-Velázquez \inst{\ref{UMSNH}} \and D.~Avila Rojas \inst{\ref{IF-UNAM}} \and H.A.~Ayala Solares \inst{\ref{PSU}} \and R.~Babu \inst{\ref{MSU}} \and E.~Belmont-Moreno \inst{\ref{IF-UNAM}} \and A.~Bernal \inst{\ref{IA-UNAM}} \and K.S.~Caballero-Mora \inst{\ref{UNACH}} \and T.~Capistrán \inst{\ref{IA-UNAM}} \and A.~Carramiñana \inst{\ref{INAOE}} \and S.~Casanova \inst{\ref{IFJ-PAN}} \and U.~Cotti \inst{\ref{UMSNH}} \and J.~Cotzomi \inst{\ref{FCFM-BUAP}} \and S.~Coutiño de León \inst{\ref{UW-Madison}} \and E.~De la Fuente \inst{\ref{UdG}} \and C.~de León \inst{\ref{UMSNH}} \and D.~Depaoli \inst{\ref{MPIK}} \and P.~Desiati \inst{\ref{UW-Madison}} \and N.~Di Lalla \inst{\ref{Stanford}} \and R.~Diaz Hernandez \inst{\ref{INAOE}} \and B.L.~Dingus \inst{\ref{LANL}} \and M.A.~DuVernois \inst{\ref{UW-Madison}} \and K.~Engel \inst{\ref{UMD}} \and T.~Ergin \inst{\ref{MSU}} \and C.~Espinoza \inst{\ref{IF-UNAM}} \and K.L.~Fan \inst{\ref{UMD}} \and K.~Fang \inst{\ref{UW-Madison}} \and N.~Fraija \inst{\ref{IA-UNAM}} \and S.~Fraija \inst{\ref{IA-UNAM}} \and J.A.~García-González \inst{\ref{ITESM}} \and F.~Garfias \inst{\ref{IA-UNAM}} \and M.M.~González \inst{\ref{IA-UNAM}} \and J.A.~Goodman \inst{\ref{UMD}} \and S.~Groetsch \inst{\ref{MTU}} \and J.P.~Harding \inst{\ref{LANL}} \and S.~Hernández-Cadena \inst{\ref{SJTU}} \and I.~Herzog \inst{\ref{MSU}} \and J.~Hinton \inst{\ref{MPIK}} \and D.~Huang \inst{\ref{UMD}} \and F.~Hueyotl-Zahuantitla \inst{\ref{UNACH}} \and T.B.~Humensky \inst{\ref{GSFC}} \and P.~Hüntemeyer \inst{\ref{MTU}} \and S.~Kaufmann \inst{\ref{UPP}} \and D.~Kieda \inst{\ref{University of Utah}} \and W.H.~Lee \inst{\ref{IA-UNAM}} \and J.~Lee \inst{\ref{UOS}} \and H.~León Vargas \inst{\ref{IF-UNAM}} \and J.T.~Linnemann \inst{\ref{MSU}} \and A.L.~Longinotti \inst{\ref{IA-UNAM}} \and G.~Luis-Raya \inst{\ref{UPP}} \and K.~Malone \inst{\ref{LANL}} \and O.~Martinez \inst{\ref{FCFM-BUAP}} \and J.~Martínez-Castro \inst{\ref{CIC-IPN}} \and J.A.~Matthews \inst{\ref{UNM}} \and P.~Miranda-Romagnoli \inst{\ref{UAEH}} \and J.A.~Montes \inst{\ref{IA-UNAM}} \and E.~Moreno \inst{\ref{FCFM-BUAP}} \and M.~Mostafá \inst{\ref{Temple}} \and M.~Najafi \inst{\ref{MTU}} \and L.~Nellen \inst{\ref{ICN-UNAM}} \and M.U.~Nisa \inst{\ref{MSU}} \and L.~Olivera-Nieto \inst{\ref{MPIK}} \and N.~Omodei \inst{\ref{Stanford}} \and Y.~Pérez Araujo \inst{\ref{IF-UNAM}} \and E.G.~Pérez-Pérez \inst{\ref{UPP}} \and C.D.~Rho \inst{\ref{SKKU}} \and D.~Rosa-González \inst{\ref{INAOE}} \and H.~Salazar \inst{\ref{FCFM-BUAP}} \and D.~Salazar-Gallegos \inst{\ref{MSU}} \and A.~Sandoval \inst{\ref{IF-UNAM}} \and M.~Schneider \inst{\ref{UMD}} \and J.~Serna-Franco \inst{\ref{IF-UNAM}} \and A.J.~Smith \inst{\ref{UMD}} \and Y.~Son \inst{\ref{UOS}} \and R.W.~Springer \inst{\ref{University of Utah}} \and O.~Tibolla \inst{\ref{UPP}} \and K.~Tollefson \inst{\ref{MSU}} \and I.~Torres \inst{\ref{INAOE}} \and R.~Torres-Escobedo \inst{\ref{SJTU}} \and R.~Turner \inst{\ref{MTU}}{\large *} \and F.~Ureña-Mena \inst{\ref{INAOE}} \and E.~Varela \inst{\ref{FCFM-BUAP}} \and L.~Villaseñor \inst{\ref{FCFM-BUAP}} \and X.~Wang \inst{\ref{MTU}} \and Z.~Wang \inst{\ref{UMD}} \and I.J.~Watson \inst{\ref{UOS}} \and E.~Willox \inst{\ref{UMD}} \and S.~Yu \inst{\ref{PSU}} \and S.~Yun-Cárcamo \inst{\ref{UMD}} \and H.~Zhou \inst{\ref{SJTU}}}

\institute{ Instituto de F\'{i}sica, Universidad Nacional Autónoma de México, Ciudad de Mexico, Mexico \label{IF-UNAM} \and  Universidad Autónoma de Chiapas, Tuxtla Gutiérrez, Chiapas, México \label{UNACH} \and  Universidad Michoacana de San Nicolás de Hidalgo, Morelia, Mexico  \label{UMSNH} \and  Department of Physics, Pennsylvania State University, University Park, PA, USA  \label{PSU} \and  Department of Physics and Astronomy, Michigan State University, East Lansing, MI, USA  \label{MSU} \and  Instituto de Astronom\'{i}a, Universidad Nacional Autónoma de México, Ciudad de Mexico, Mexico \label{IA-UNAM} \and  Instituto Nacional de Astrof\'{i}sica, Óptica y Electrónica, Puebla, Mexico \label{INAOE} \and  Institute of Nuclear Physics Polish Academy of Sciences, PL-31342 IFJ-PAN, Krakow, Poland  \label{IFJ-PAN} \and  Facultad de Ciencias F\'{i}sico Matemáticas, Benemérita Universidad Autónoma de Puebla, Puebla, Mexico \label{FCFM-BUAP} \and  Department of Physics, University of Wisconsin-Madison, Madison, WI, USA  \label{UW-Madison} \and  Departamento de F\'{i}sica, Centro Universitario de Ciencias Exactase Ingenierias, Universidad de Guadalajara, Guadalajara, Mexico \label{UdG} \and  Max-Planck Institute for Nuclear Physics, 69117 Heidelberg, Germany \label{MPIK} \and  Department of Physics, Stanford University: Stanford, CA 94305–4060, USA \label{Stanford} \and  Los Alamos National Laboratory, Los Alamos, NM, USA  \label{LANL} \and  Department of Physics, University of Maryland, College Park, MD, USA  \label{UMD} \and  Tecnologico de Monterrey, Escuela de Ingenier\'{i}a y Ciencias, Ave. Eugenio Garza Sada 2501, Monterrey, N.L., Mexico, 64849 \label{ITESM} \and  Department of Physics, Michigan Technological University, Houghton, MI, USA  \label{MTU} \and  Universidad Politecnica de Pachuca, Pachuca, Hgo, Mexico  \label{UPP} \and  University of Seoul, Seoul, Rep. of Korea \label{UOS} \and  Centro de Investigaci\'{o}n en Computaci\'{o}n, Instituto Polit\'{e}cnico Nacional, M\'{e}xico City, M\'exico. \label{CIC-IPN} \and  Dept of Physics and Astronomy, University of New Mexico, Albuquerque, NM, USA  \label{UNM} \and  Universidad Autónoma del Estado de Hidalgo, Pachuca, Mexico  \label{UAEH} \and  Instituto de Ciencias Nucleares, Universidad Nacional Autónoma de Mexico, Ciudad de Mexico, Mexico  \label{ICN-UNAM} \and  Department of Physics, Sungkyunkwan University, Suwon 16419, South Korea \label{SKKU} \and  Department of Physics and Astronomy, University of Utah, Salt Lake City, UT, USA  \label{University of Utah} \and  Tsung-Dao Lee Institute \& School of Physics and Astronomy, Shanghai Jiao Tong University, Shanghai, China \label{SJTU} \and  Department of Physics, Temple University, Philadelphia, Pennsylvania, USA \label{Temple} \and NASA Goddard Space Flight Center, Greenbelt, MD, USA \label{GSFC}}

\offprints{Rhiannon Turner,
\protect\\\email{rturner1@mtu.edu};
\protect\\\protect * Corresponding author
}

   \date{Received July 15, 2024; Accepted Sept. 9, 2024}

 
  \abstract
   {Supernova remnants (SNRs) are believed to be capable of accelerating cosmic rays (CRs) to PeV energies. SNR G106.3+2.7 is a prime PeVatron candidate. It is formed by a head region, where the pulsar J2229+6114 and its boomerang-shaped pulsar wind nebula are located, and a tail region containing SN ejecta. The lack of observed gamma ray emission from the two regions of this SNR has made it difficult to assess which region would be responsible for the PeV CRs.}
   {We aim to characterize the very-high-energy (VHE, 0.1--100 TeV) gamma ray emission from SNR G106.3+2.7 by determining the morphology and spectral energy distribution of the region. This is accomplished using 2565 days of data and improved reconstruction algorithms from the High Altitude Water Cherenkov (HAWC) Observatory. We also explore possible gamma ray production mechanisms for different energy ranges.}
   {Using a multi-source fitting procedure based on a maximum-likelihood estimation method, we evaluate the complex nature of this region. We determine the morphology, spectrum, and energy range for the source found in the region. Molecular cloud information is also used to create a template and evaluate the HAWC gamma ray spectral properties at ultra-high-energies (UHE, $>$56 TeV). This will help probe the hadronic nature of the highest-energy emission from the region.}
   {We resolve one extended source coincident with all other gamma ray observations of the region. The emission reaches above 100~TeV and its preferred log-parabola shape in the spectrum shows a flux peak in the TeV range. The molecular cloud template fit on the higher energy data reveals that the SNR's energy budget is fully capable of producing a purely hadronic source for UHE gamma rays.}
   {The HAWC observatory resolves one extended source between the head and the tail of SNR G106.3+2.7 in the VHE gamma ray regime. The template fit suggests the highest energy gamma rays could come from a hadronic origin. However, the leptonic scenario, or a combination of the two, cannot be excluded at this time.}


   \maketitle
%

\section{Introduction}
\label{sec:intro}
   
Cosmic-ray (CR) accelerators have been probed indirectly across a wide range of energies, from the radio to ultra-high-energy gamma rays. Galactic CRs are expected to be accelerated up to the knee of the CR spectrum, meaning they can reach up to PeV energies \citep{cristofari}. It is suggested that these PeV CR accelerators (PeVatrons) are most commonly supernova remnants (SNRs), where charged particles are accelerated in the shock fronts \citep{Cristofari_2018, Reynolds_2010, Bell_2013}.  
   
SNR G106.3+2.7 is a PeVatron candidate that has been under study for many years now. It was first detected in 1990 by \cite{j&h} at 408~MHz with the Dominion Radio Astrophysical Observatory (DRAO). The head of the SNR is home to the pulsar J2229+6114 and its boomerang-like pulsar wind nebula (PWN) \citep{kothes2001}, known as the Boomerang region. An extended shell then reaches into the tail region of the SNR \citep{p&j}. There have been many proposed distances for this region. \cite{p&j} adopted a distance of 12~kpc for the SNR based on a flat rotation curve and \cite{Halpern_2001} found the distance to be 3~kpc using X-ray absorption column density. HI observations from \cite{kothes2001} place the system at 800 PC. 

   \begin{figure*}
   \sidecaption
   \includegraphics[width=12cm]{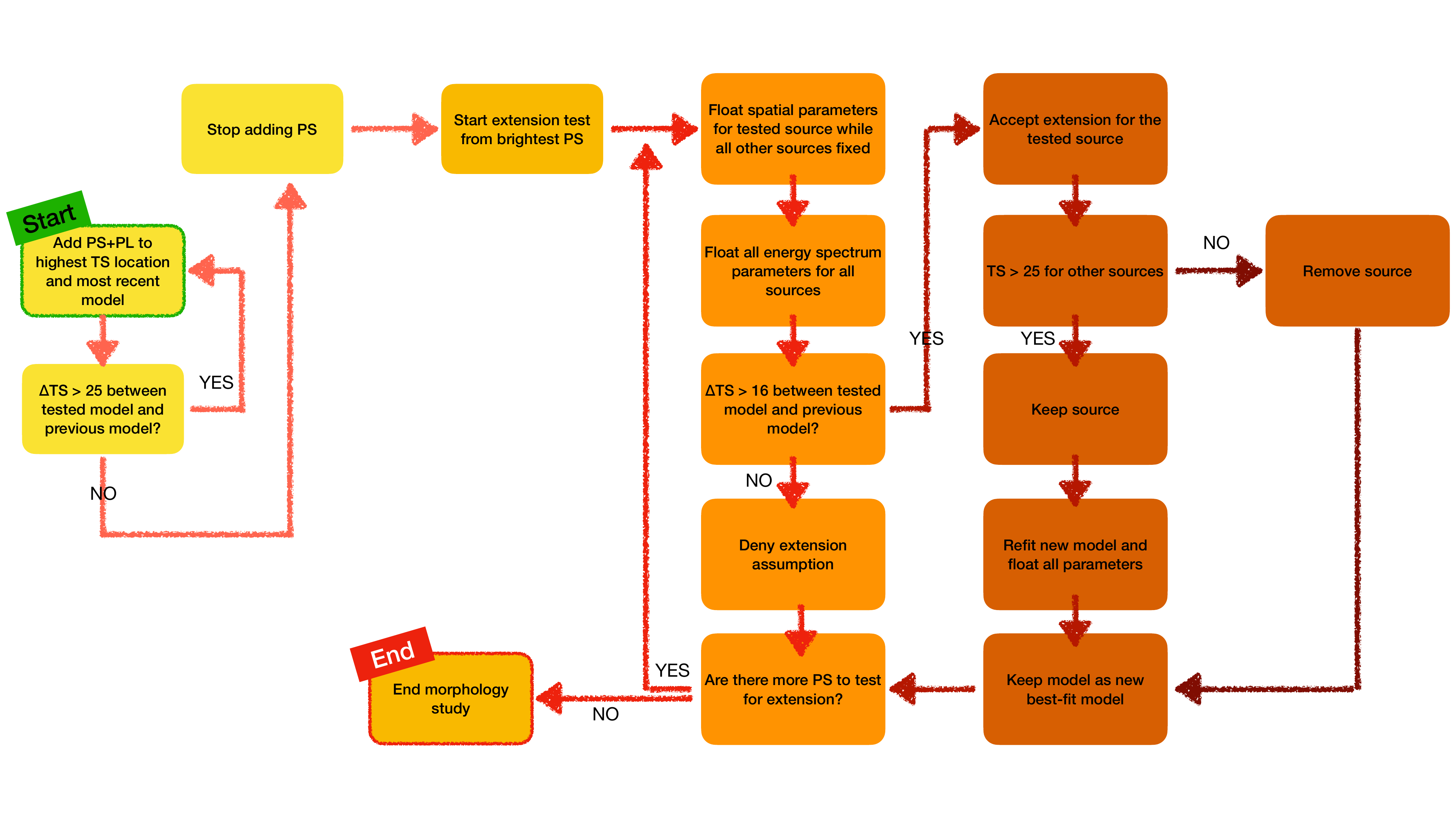}
   \caption{Diagram of the multi-source search method described in section~\ref{sec:analysis}. Yellow boxes (first two columns) describe how point sources (PSs) are added, where PL refers to a power-law spectrum. Orange/red boxes (last four columns) describe how extended sources are added.}
              \label{fig:systematic_diagram}
    \end{figure*}
   
X-ray observations show similar structures coincident with the head and tail regions. \cite{fujita} finds that the X-ray emission from the entire SNR is from nonthermal synchrotron radiation. However, the X-ray brightness in the head region contrasts with the gamma ray observations, which instead see a brightness in the tail region.  

Ground-based gamma ray particle detector arrays have so far only been able to resolve one emission spot in the Boomerang region. Milagro, VERITAS, Tibet, and LHAASO report extended emission coincident with the tail of the SNR. \textit{Fermi}-LAT \citep{fermi} and MAGIC \citep{magic} report gamma ray emission from the head and tail regions of the SNR separately, but no other gamma ray observations have been used to disentangle the sources thus far. \cite{fermi} separates the pulsar emission from the SNR and performs a detailed analysis of the tail region using \textit{Fermi}-LAT data. They find that pion decay (PD) is the primary mechanism for producing the high-energy emission seen in the tail region. 
   
\cite{magic} reports 0.16$^{\circ}$ extended gamma ray emission from the head and tail regions. They also show that the emission shifted in the SNR depending on the energy threshold. Their highest energy emission (6.0-30~TeV) is consistent with the tail of the SNR. Their multi-wavelength analysis reveals that the head region can have a leptonic or hadronic explanation, while the tail region is most likely of hadronic origin. 

The previous HAWC detection analyzes 1350 days of data using older reconstruction algorithms \citep{hawc}. \cite{hawc} finds point-like emission consistent with the other gamma ray observations, which see extended emission in the tail region. The work presented in this paper uses more data and improved reconstruction algorithms \citep{pass5}, which further extends the previous HAWC measurement past 110 TeV with flux points. We also explore a molecular cloud template fit using data from the \textit{Planck} 353 GHz dust opacity map and estimate the CR flux in the region using distance information from the Dame CO Survey \citep{Dame_2001}. Section~\ref{sec:hawc} describes the HAWC observatory and data used for this analysis. Section~\ref{sec:analysis} describes the analysis procedure, as well as the results from the procedure. Gamma ray production mechanisms and their probable CR sources are described in section~\ref{sec:GR-production}. Conclusions and further outlooks are presented in section~\ref{sec:conclusion}. 


\section{HAWC Observatory and data}
\label{sec:hawc}

   The High-Altitude Water Cherenkov (HAWC) Observatory is located in Puebla, Mexico at an altitude of 4100 m. HAWC utilizes the water Cherenkov detection technique and consists of 300 water Cherenkov detectors (WCDs). Each WCD contains four photo-multiplier tubes (PMTs) and the entire array covers $\sim$22000~m\textsuperscript{2} \citep{nim}. For this analysis, we used 2565 days of data, which is about 1000 days more data than the previous publication \cite{hawc}. A neural net (NN) algorithm was used to reconstruct the energy of the primary gamma ray that initiated the extensive air shower. From here, the data was binned in a 2D binning scheme based on the fraction of PMTs triggered and reconstructed energy \citep{crab2019}. In addition to having more data, the new data reconstruction process used algorithms that better optimize angular resolution. 


\section{Analysis and results}
\label{sec:analysis}
\subsection{Systematic source search}

\begin{figure}
\centering
\subfigure[]{\includegraphics[width=5.9cm]{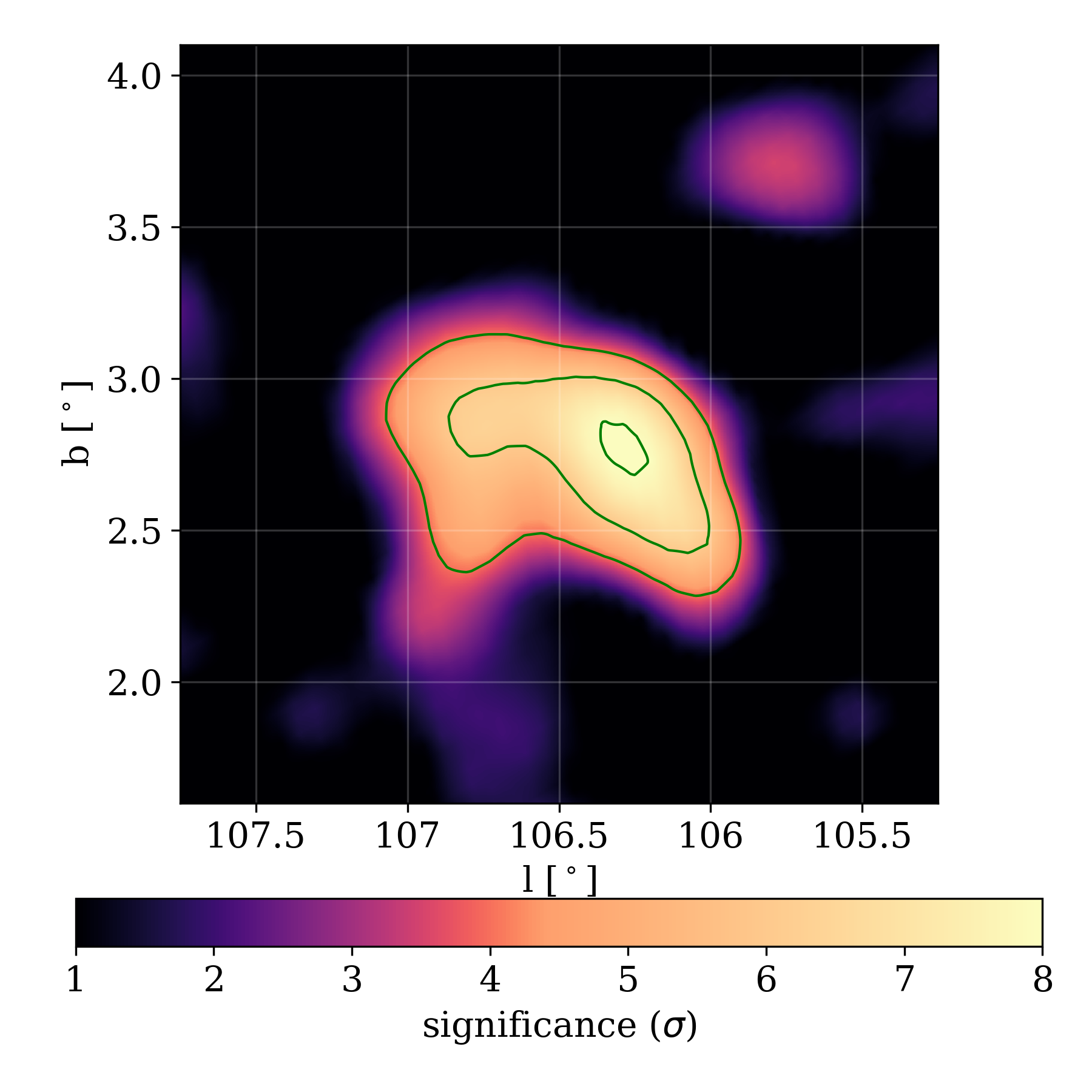}}\\
\subfigure[]{\includegraphics[width=5.9cm]{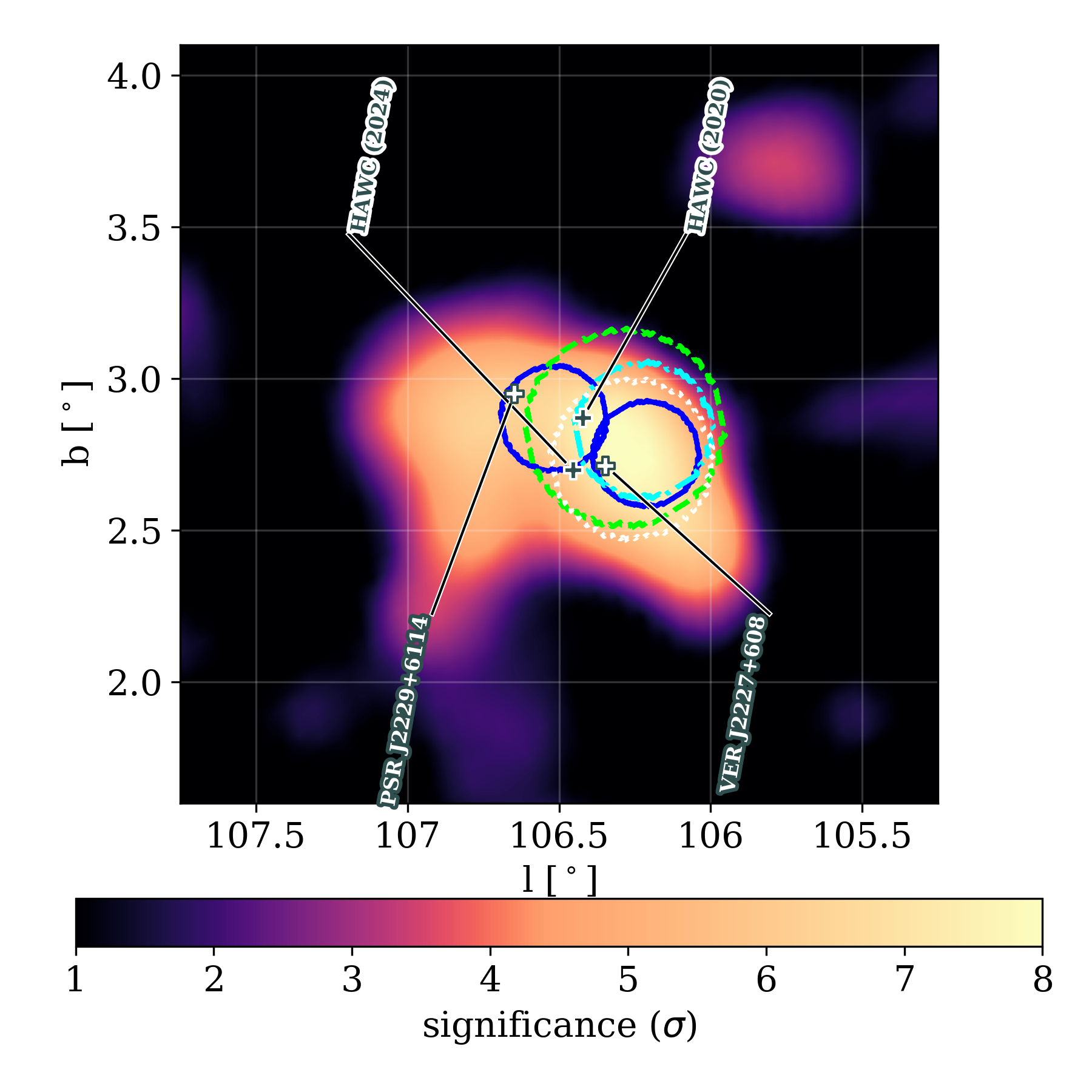}}\\
\subfigure[]{\includegraphics[width=5.9cm]{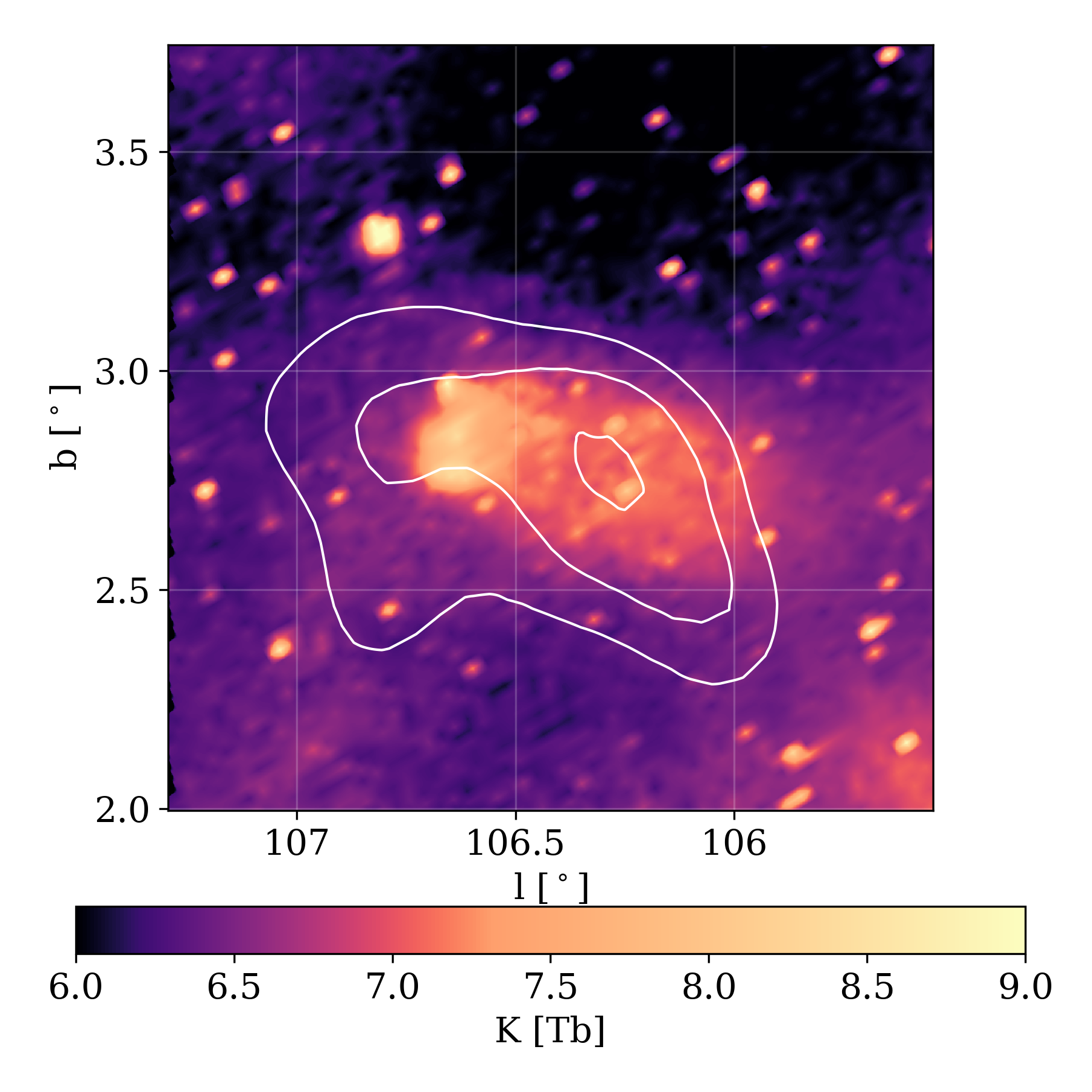}}
\caption{Significance maps for HAWC's all-energy emission and its comparison to the SNR in the region. \textbf{a)} The HAWC all-energy significance map of the region with the 4, 6, and 8$\sigma$ contours overlaid in green. \textbf{b)} The HAWC significance map of the region with labels showing the best-fit positions from this analysis, the previous HAWC publication \citep{hawc}, and VERITAS \citep{veritas}. The green dashed, white dotted, cyan dot-dashed, and blue solid contour indicate LHAASO's analysis region \citep{lhaaso}, Tibet AS$\gamma$'s 1$\sigma$ extension \citep{tibet}, \textit{Fermi}-LAT's 1$\sigma$ extension \citep{fermi}, and MAGIC's analysis regions \citep{magic}, respectively. \textbf{c)} The brightness temperature image \citep{cgps} of the SNR overlaid with the HAWC 4, 6, and 8$\sigma$ contours in white. The bright emission on the left shows the head of the SNR, while the faint emission extending to the right shows the tail. }
\label{fig:region_maps}
\end{figure}

This first part of this analysis utilized the Multi-Mission Maximum Likelihood (threeML) framework \citep{threeml}, along with the HAWC Accelerated Likelihood (HAL) plugin \citep{hal} to work through a step-by-step multi-source search method inspired by the \textit{Fermi}-LAT Extended Source Catalog \citep{FGES, 4hwc}. ThreeML uses a likelihood ratio as a test statistic (TS) to reach a parameter set that maximizes the likelihood of the hypothesized model over the background-only model. The value of the TS is defined as follows: 

\begin{equation}
\label{eq:TS}
TS = 2\ln{\frac{L_{\mathrm{alt}}}{L_{\mathrm{null}}}},
\end{equation}

\noindent  
where L$_{alt}$ is the model hypothesis and L$_{null}$ is the background only hypothesis. In order to arrive at the most statistically preferred model in our multi-source search method, we looked at the difference in TS between each nested model that is tested. There are two main components to this analysis described below: point source search and extension testing. 

\noindent
\underline{Point Source Search} -- In this section, source TS is calculated assuming L$_{alt}$ to be the current model and L$_{null}$ to be the previous model.
  \begin{enumerate}
   \item The first step of our multi-source search is to add one point source with a power-law (PL) spectrum to the location of the highest TS value in the significance map, leaving the position and spectral values of the newly added source free while the position of the previous sources remains fixed.
   \item After fitting the model in step 1, we check the $\Delta$TS between the previous model and the most recently tested model:
    \begin{enumerate}
      \item If the $\Delta$TS~$>$~25, we keep the source in the model and return to step 1.
      \item If the $\Delta$TS~$<$~25, we do not add the additional source to the model and stop adding point sources. We then start our extended source testing.
     \end{enumerate}
   \end{enumerate}
\underline{Extended Source Testing} -- In this section, source TS is calculated assuming L$_{alt}$ to be the entire model and L$_{null}$ to be the entire model with the source in question subtracted out.
  \begin{enumerate}
   \item We start by replacing the highest TS point source with a 2D extended Gaussian and leave all other sources as point sources. We keep all source locations fixed.
   \item After fitting the model in step 1, we check the $\Delta$TS between the previous model and the most recently tested model:
     \begin{enumerate}
      \item If the $\Delta$TS~$>$~16, we accept the extended source model and go to step 3.
      \item If the $\Delta$TS~$<$~16, we reject the extended source model and move to the next highest TS value source. We then start back at step 1.
     \end{enumerate}
   \item Next, we check the TS values of the point sources in the model:
   \begin{enumerate}
      \item If the TS values of all point sources are $>$~25, we refit the model and float all parameters. If untested point sources remain, we go back to step 1. We end the study if there are no more point sources remaining for extension testing.
      \item If the TS of any point source is $<$~25, we remove the source(s) and refit the new model while floating all parameters. If untested point sources remain, we go back to step 1. We end the study if there are no more point sources remaining for extension testing.
     \end{enumerate}
 \end{enumerate}

 Once all point sources in the final point-source model are tested for extension, the source search ends. Figure~\ref{fig:systematic_diagram} visualizes the point and extended source systematic steps. After completing the source search, we then test each source for a curvature in its spectrum using a log-parabola (LP), which is defined by:

\begin{equation}
\label{eq:PL}
\frac{dN}{dE}=N_{0}\bigg( \frac{E}{30 \mathrm{TeV}} \bigg) ^{\alpha -\beta \log{\left( \frac{E}{30 \mathrm{TeV}} \right)}},
\end{equation}
\noindent
where \textit{N$_0$} is the normalization parameter, the pivot energy is 30~TeV, $\alpha$ is the spectral index, and $\beta$ dictates the curve in the spectrum. Since the simple power-law is nested within the LP, we can use the same TS comparison to decide if a source's spectrum is curved or not. We start the curvature testing with the brightest source. If the new model (with the curved spectrum) has a $\Delta$TS~>~16, we keep it and move on to test the next source. If the new model (with the curved spectrum) has a $\Delta$TS~<~16, it is rejected and we test the next source. Once all sources have been tested for curvature, we refit the final model to ensure the best-fit parameters are further optimized.
 
 After completing this procedure in the Boomerang region, the best-fitting model was found to be a single extended source with an LP spectrum. Table~\ref{tab:best-fit_values} lists the best-fit position and spectral parameters with their statistical and systematic uncertainties. The HAWC significance maps using the new reconstruction algorithms and best-fit values are shown in figure~\ref{fig:region_maps}. The systematic uncertainties were determined using various detector response files that test possible effects on the data resulting from differences in the modeling of the instrument. These differences are calculated following \cite{crab2019}. 
 
 The next step is to determine the energy range in which the source is confidently detected. The energy range was calculated using the method described in \cite{crab2017}. The energy range at a 1$\sigma$ confidence level, which corresponds to a 68\% confidence limit, is 4--145~TeV. The flux points were calculated by first binning the data into energy bins and determining the median energy in a given bin. We then used threeML to fit the normalization parameter \textit{N$_0$} at the median energy (E) in equation~\ref{eq:PL} and all other parameters are fixed at their best-fit values from table~\ref{tab:best-fit_values} \citep{crab2019}.
 
\begin{figure}
   \centering
   \resizebox{\hsize}{!}{\includegraphics{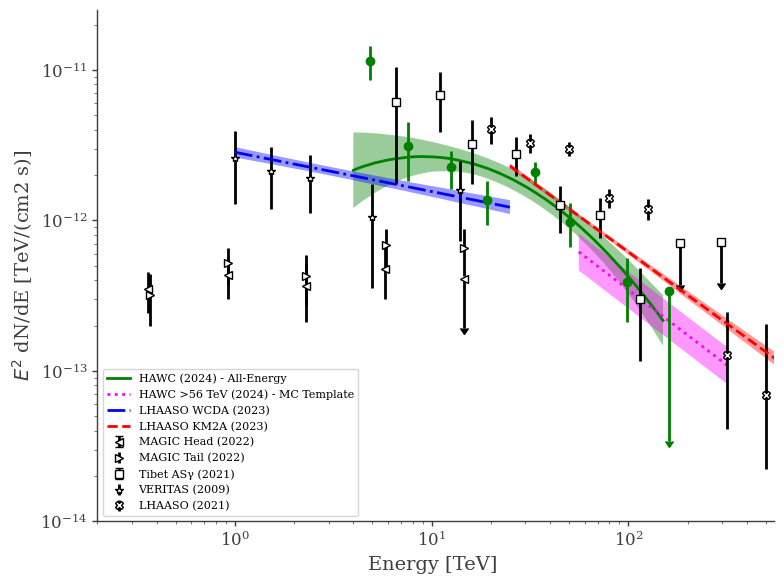}}
   \caption{Spectral energy distribution (SED) of the ground-based VHE gamma ray observations. The green SED with circle flux points are the all-energy HAWC results from this work. The magenta dotted line is the HAWC result for the $>$56~TeV molecular cloud template analysis. The red line (KM2A) and blue line (WCDA) are from the first LHAASO catalog \citep{lhaaso_cat}. All SEDs show only statistical uncertainties. The open stars, right-facing triangles, left-facing triangles, squares, and x-marks correspond to the VERITAS \citep{veritas}, MAGIC tail \citep{magic}, MAGIC head \citep{magic}, Tibet AS$\gamma$ \citep{tibet}, and LHAASO \citep{lhaaso} measurements, respectively. }
              \label{fig:gr_spectrum}
    \end{figure}

The position (figure~\ref{fig:region_maps}.b) and spectrum (figure~\ref{fig:gr_spectrum}) of the new HAWC observation are consistent with the other gamma ray observations in this region. \cite{tibet} reports a source extension of 0.24$^{\pm0.1 (stat)}_{\pm0.1 (sys)}$ degrees (shown in figure~\ref{fig:region_maps}.b) and LHAASO uses an extension of 0.3 degrees in their high-energy source catalog (shown in figure~\ref{fig:region_maps}.b), as well as reports an extension of 0.35$\pm0.01$ (WCDA) and 0.25$\pm0.02$ (KM2A) degrees in their first catalog \citep{lhaaso, lhaaso_cat}. All of these measurements are within the systematic uncertainties of the extension found in this work (table~\ref{tab:best-fit_values}). The main differences between the various gamma ray observations are most likely attributed to each observatory detecting different contributions from the head and tail regions across varying energy ranges.

   \begin{table*}[h]
   \caption{Best-fit results from the source search. The first uncertainties listed are statistical and the second uncertainties are systematic. The spectral values are quoted assuming the spectral parameters as defined in equation~\ref{eq:PL}.}
   \centering 
   \label{tab:best-fit_values}      
   \begin{tabular}{cccccc} 
   \hline\hline
      \rule[-1ex]{0pt}{3.5ex} RA [$^{\circ}$] & DEC [$^{\circ}$] & Radius [$^{\circ}$]  \\
   \hline
      \rule[-1ex]{0pt}{3.5ex} 337.20 $^{(+0.12)}_{(-0.12)}$$^{(+0.26)}_{(-0.33)}$ & 60.92 $^{(+0.05)}_{(-0.05)}$$^{(+0.13)}_{(-0.11)}$ & 0.34 $^{(+0.04)}_{(-0.04)}$$^{(+0.12)}_{(-0.13)}$ \\
   \hline\hline
   \rule[-1ex]{0pt}{3.5ex} N$_{0}$ [$\times 10^{-15} \mathrm{cm\textsuperscript{-2}  TeV\textsuperscript{-1}  s\textsuperscript{-1}}$] & $\alpha$ & $\beta$ \\
   \hline
    \rule[-1ex]{0pt}{3.5ex} 1.86 $^{(+0.32)}_{(-0.27)}$$^{(+1.33)}_{(-1.10)}$ & 2.76 $^{(+0.16)}_{(-0.16)}$$^{(+0.84)}_{(-0.60)}$ & 0.32 $^{(+0.13)}_{(-0.13)}$$^{(+1.05)}_{(-0.49)}$ \\
   \end{tabular}
   \end{table*}

\subsection{HAWC data $>$56 TeV}

We will now explore the possible production mechanisms that would contribute to the PeVatron's UHE emission in this region. Most PeVatron theory assumes that these systems are hadronic, which is confirmed by correlation with molecular clouds in the area. However, recent studies suggest that the leptonic scenario is more than capable of producing UHE emission as well \citep{Vannoni, Breuhaus}. Both scenarios will be considered for this region.

\subsubsection{Molecular cloud template fitting}
\label{sec:MC_fit}

We will first look at the UHE gamma ray emission in the region using the pion decay production mechanism. The head of the SNR houses the pulsar and its PWN, which would mean that the protons are most likely being accelerated in the tail region where there is supernova (SN) ejecta and the shock front, which is coincident with a nearby molecular cloud. The molecular cloud is an ideal target for the accelerated protons to interact with and undergo PD. This means that UHE gamma ray morphology should take the shape of the molecular clouds that it is being produced in. This analysis explores fitting HAWC data above 56~TeV using a molecular cloud template from the \textit{Planck} Collaboration \citep{Planck} to explain a possible hadronic mechanism. 

\begin{figure*}
\sidecaption
  \includegraphics[width=12cm]{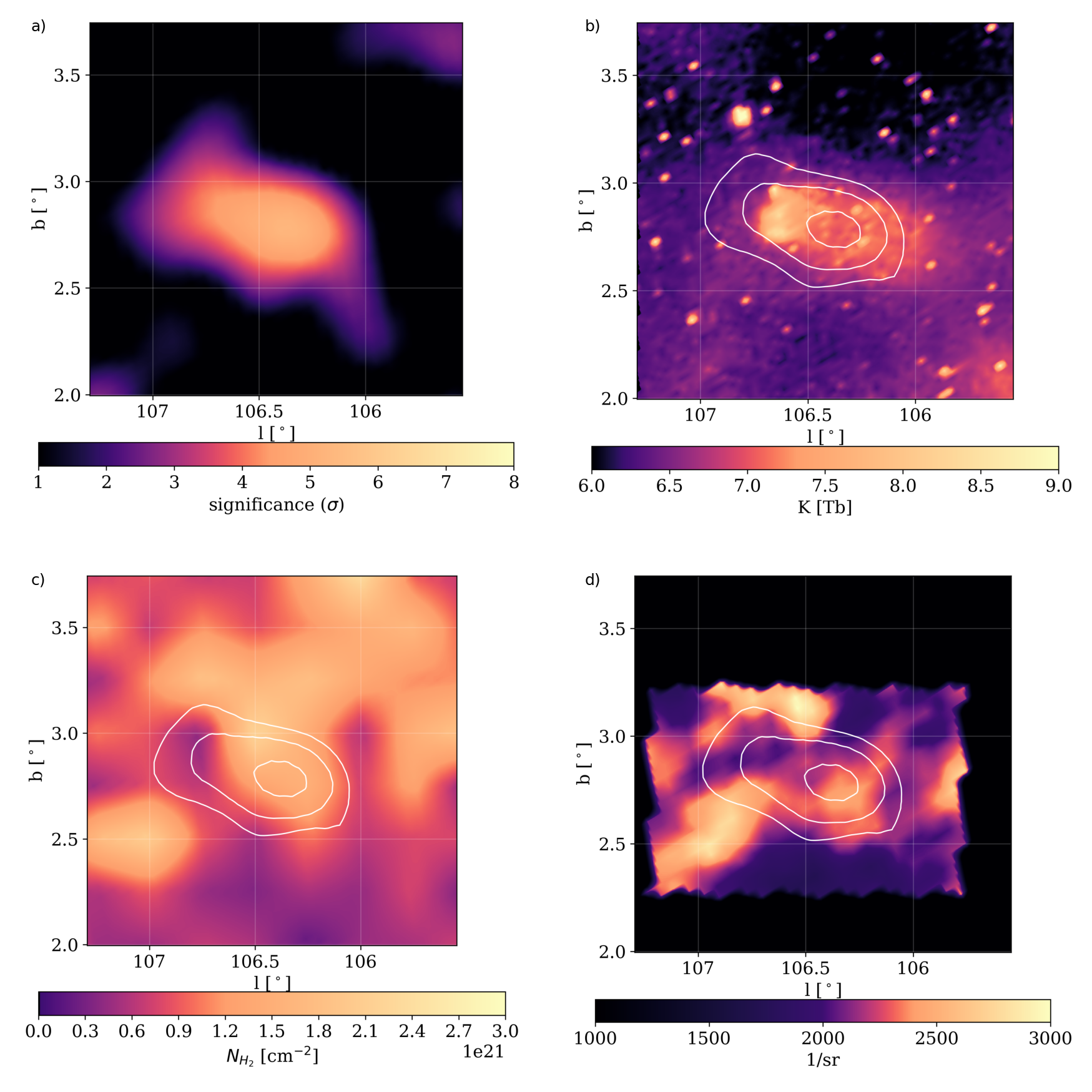}
  \caption{Significance maps for HAWC's $>$56~TeV emission and its comparison to the SNR and molecular clouds in the region. \textbf{a)} The HAWC significance map of the region above 56~TeV. \textbf{b)} The brightness temperature image \citep{cgps} of the SNR overlaid with the HAWC 3, 4, and 5$\sigma$ contours in white. \textbf{c)} Molecular hydrogen column density integrated over a velocity range of $-20$~km/s to 0~km/s \citep{Dame_2001}. \textbf{d)} The \textit{Planck} 353 GHz template normalized to 1/sr that is used in the UHE HAWC fit. }
  \label{fig:vhe_maps}
\end{figure*}

We used two different surveys to assess the molecular clouds in the region: the Dame CO survey \citep{Dame_2001} and the 353 GHz \textit{Planck} dust opacity map \citep{Planck}. Figure~\ref{fig:vhe_maps}.c shows significant CO emission from the Dame CO survey coincident with HAWC's $>$56 TeV 3, 4, and 5$\sigma$ contours. Ultimately, we used the \textit{Planck} dust map for the template fitting because it is more recent than the Dame survey (Dame 2001 vs. \textit{Planck} 2011) and provides information for HI gas that is optically thick in CO. However, the \textit{Planck} dust map has no distance information, so we use the Dame CO survey to confirm that there is no emission in front of or behind the Boomerang region that is being included in the \textit{Planck} dust map. Appendix~\ref{sec:validation} gives a more in-depth explanation of how the \textit{Planck} dust map was validated for use. 

Again, we utilized the threeML/HAL framework to do a maximum likelihood fitting of the molecular cloud spatial template. A similar procedure to that of \cite{gmc_hawc} was used to create the spatial template that is fed to threeML for the template fitting:

\begin{enumerate}
    \item We start by calculating the column density-- this calculation is taken from \cite{gmc_hawc}:
\begin{equation}
    N_{H} = \tau_{D} / \bigg(\frac{\tau_{D}}{N_{H}}\bigg)^{ref},
\end{equation}
where ($\tau_{D}/N_{H}$)$^{ref}$ = (1.18~$\pm$~0.17)~$\times$~10$^{-26}$~cm$^2$ for the 353~GHz \textit{Planck} map \citep{Planck} and $\tau_{D}$ is the dust opacity.
    \item We then crop the data to the size of the region of interest.
    \item We calculate the mass of the region using the following equation:
\begin{equation}
    M_{dust} = N_{H} \Omega d^{2} m_{h},
\end{equation}
where $\Omega$ is the angular area of the cloud and $d~=~800~pc$ as the distance to the cloud \citep{kothes2001}
    \item Finally, we normalize the data to 1/sr. This is required for a 3ML 2D template. 
\end{enumerate}

Since \cite{kothes2001} used HI observations to determine the distance to the region, we adopted the same distance in this paper for our calculations. Figure~\ref{fig:vhe_maps}.d shows the \textit{Planck} dust map template with HAWC's $>$56 TeV 3, 4, and 5$\sigma$ contours. Appendix~\ref{sec:template} gives additional information and important caveats for the template itself. This template, along with a power-law spectrum and a fixed index of $\mathrm{-3.0}$, was used to fit HAWC's UHE emission shown in figure~\ref{fig:vhe_maps}.a. The spectral index for this fit was chosen to be slightly softer than the all-energy measured index ($\mathrm{-2.76}$) as a way to capture the curvature that causes a sharp dropoff at higher energies (figure~\ref{fig:gr_spectrum}). The best-fit normalization at 50 TeV for the template fit is 2.8~$^{+0.9}_{-0.7}$~$\times$~10$^{-16}$~($\mathrm{1/TeV~s~cm^{2}}$). 

\subsubsection{Simple leptonic model fitting}
\label{sec:simple_fit}

We also explore the possibility of a leptonic mechanism producing the UHE HAWC data. This would mean that electrons are being accelerated by the SN ejecta or the PWN winds. For this scenario, we modeled the region with a point source morphology and a power-law spectrum, which has a fixed index of $-3.0$. Table~\ref{tab:vhe_best-fit} lists the best-fit values for the position and flux normalization after fitting the emission shown in figure~\ref{fig:vhe_maps}.a with the simple model.

   \begin{table*}[h]
   \caption{Best-fit results from the simple modeling above 56 TeV, which assumes a point source with a PL spectrum. All errors are statistical.}
   \centering 
   \label{tab:vhe_best-fit}      
   \begin{tabular}{cccccc} 
   \hline\hline
      \rule[-1ex]{0pt}{3.5ex}  RA [$^{\circ}$] & DEC [$^{\circ}$] & N$_{0}$ [$\mathrm{cm\textsuperscript{-2}  TeV\textsuperscript{-1}  s\textsuperscript{-1}}$] & Index & Pivot [TeV] & TS  \\
   \hline
      \rule[-1ex]{0pt}{3.5ex} 337.05 $\pm$ 0.07 & 60.92 $\pm$ 0.04 & 1.7$^{+0.6}_{-0.4}$ $\times$ 10$^{-16}$ & 3.0 & 50 & 29 \\
   \end{tabular}
   \end{table*}
   

\section{Scenarios for gamma ray production}
\label{sec:GR-production}
\subsection{Cosmic ray energy density from protons}

In section~\ref{sec:MC_fit}, we explored a hadronic-related fit to the HAWC UHE data with a molecular cloud template. We estimated the CR energy density using purely hadronic interactions and the measured gamma ray flux. The CR energy density can then be used to roughly determine the amount of energy that would be needed for the CR population in the region. We started by calculating the gamma ray energy flux using the best-fit gamma ray spectrum from our molecular cloud template fit:

\begin{equation}
    J = \int_{Ei}^{Ef} E \frac{dN}{dE} dE,
\end{equation}

\noindent
where 

\begin{equation}
    \frac{dN}{dE} = k \bigg( \frac{E}{E_{\mathrm{piv}}} \bigg)^{-\alpha},
\end{equation}

\noindent
and $E_i$~=~56~TeV, $E_f$~=~316~TeV, and $k$, $E_{piv}$ and $\alpha$ are the same best-fit values from the fitting in section~\ref{sec:MC_fit}. Next, we used the integrated flux to calculate the luminosity:

\begin{equation}
    L(\geq E_\gamma) = 4 \pi d^2 J,
\end{equation}

\noindent
where d~=~800~pc. Finally, we used the luminosity to calculate the CR proton energy density using the same approach as \cite{hess_nature}:

\begin{equation}
    \omega_{CR}(\geq 10E_\gamma) \approx 1.8 \times 10^{-2} \bigg( \frac{\eta_N}{1.5} \bigg)^{-1} \bigg( \frac{L(\geq E_\gamma)}{10^{34} \mathrm{erg/s}} \bigg) \bigg( \frac{M}{10^6 M_{\odot}} \bigg)^{-1},
\end{equation}

\noindent
where $\eta_N$~=~1.5 for heavier nuclei, $M~=~0.23~\times~10^5 M_{\odot}$ and $\omega_{CR}$ is in $\mathrm{eV/cm^{3}}$, which is calculated the same way as in \cite{gmc_hawc}. Here, we assumed that the proton energy scales as 10$\times$ that of the gamma ray energy \citep{Cristofari_2018}, so we are probing protons that are $>$560 TeV. The CR density was found to be 10.3~$\times$~10$^{-3}$~$\mathrm{eV/cm^{3}}$ for protons $>$560 TeV. 

The SNR has a length of 14~pc and a width of 6~pc \citep{kothes2001}. Since the tail is more elongated than the head, we assumed that the tail is two-thirds of the length of the SNR. We also assumed that the SNR is also capable of expanding 14~pc in all other directions as well, so we use a cube of 6~pc~$\times$~9.33~pc~$\times$~28~pc as the volume of the tail region as a higher estimate of the energy budget. Using these dimensions and the CR energy density, we found that the energy budget for these CR protons that are $>$560 TeV is 7.62~$\times$~10$^{44}$~erg. 

The hadronic-type modeling in this section assumes that protons are producing the UHE HAWC data. After being accelerated in the shock fronts of the SNR ejecta, the protons travel to the nearby molecular cloud and produce gamma rays through PD. We get a proton energy budget of 7.62~$\times$~10$^{44}$~erg for protons $>$560 TeV using the results from the molecular cloud template fit. This is well below the SNR's lower-end energy budget estimation of 7~$\times$~10$^{49}$~ergs \citep{kothes2001}, so the SNR is fully capable of producing the UHE gamma rays we are seeing from HAWC. Following the logic from \cite{tibet}, this SNR, which is capable of accelerating protons up to at least $\approx$0.5~PeV, should also be capable of producing protons up to $\approx$1.6~PeV during its free expansion phase at 1~kyr. Since the SNR age is believed to be 10 kyr, the diffusion of protons out to the molecular cloud may have been suppressed until now.

   \begin{table*}[h]
   \caption{Best-fit results from Naima modeling, where W$_{e}$ is the electron energy budget. All errors are statistical.}
   \centering 
   \label{tab:naima_best-fit}      
   \begin{tabular}{cccccc} 
   \hline\hline
      \rule[-1ex]{0pt}{3.5ex}  log(A [1/TeV]) & $\alpha$ & E$_{\mathrm{cutoff}}$ [TeV] & B-Field [$\mu$G] & W$_{e}$ $>$ 1 TeV [erg]  \\
   \hline
      \rule[-1ex]{0pt}{3.5ex} 41.4 $\pm$ 0.0701 & 2.57 $\pm$ 0.0237 & 208$^{+255}_{-115}$ & 4.68 $\pm$ 0.459 & 1.28$\times10^{45}$ \\
   \end{tabular}
   \end{table*}

\subsection{Multi-wavelength modeling for electrons}
\label{sec:naima}

We used the results from section~\ref{sec:simple_fit} to explore the possible leptonic nature of the UHE HAWC emission through multi-wavelength modeling. Since we assumed that the hadronic nature of the UHE gamma ray emission stems from the shock fronts in the tail of the SNR, we assumed the same scenario for the leptonic emission we model here. Therefore, we used other gamma ray observations that also modeled the tail region for additional information during our modeling. We used flux points from HAWC, MAGIC's tail source \citep{magic}, and \textit{Fermi}-LAT \citep{fermi} to model the gamma ray production using Naima \citep{naima}. The X-ray observation is from the Suzaku telescope \citep{Ge_2021} and its flux was calculated using the "East" region and a total solid angle of 17.8$^{2}$ arcmin$^{2}$. The radio observations come from the DRAO at 408 MHz and 1240 MHz \citep{kothes2001}, the Sino-German $\lambda$6 cm polarization survey \citep{sino-german}, and the Effelsberg $\lambda$11 and $\lambda$21 cm surveys \citep{sino-german}. Since the radio surveys for the Sino-German and Effelsberg surveys are reported for the entire SNR, their fluxes were scaled using the DRAO 408 MHz tail-to-whole flux ratio. 

The multi-wavelength modeling was done using the Naima framework \citep{naima}. All parent particle spectral models used a power-law with an exponential cutoff,

\begin{equation}
    f(E) = A (E / E_0) ^ {-\alpha} \exp (- (E / E_{\mathrm{cutoff}})^\beta)
\end{equation}

\noindent
where $A$ is the normalization parameter, $E_{0}$ is the pivot energy, $\alpha$ is the spectral index, $E_{cutoff}$ is the cutoff energy, and $\beta$ is the curvature parameter, set to the default value, 1.0, for all modeling. 

\begin{figure}
   \centering
   \resizebox{\hsize}{!}{\includegraphics[height=12cm]{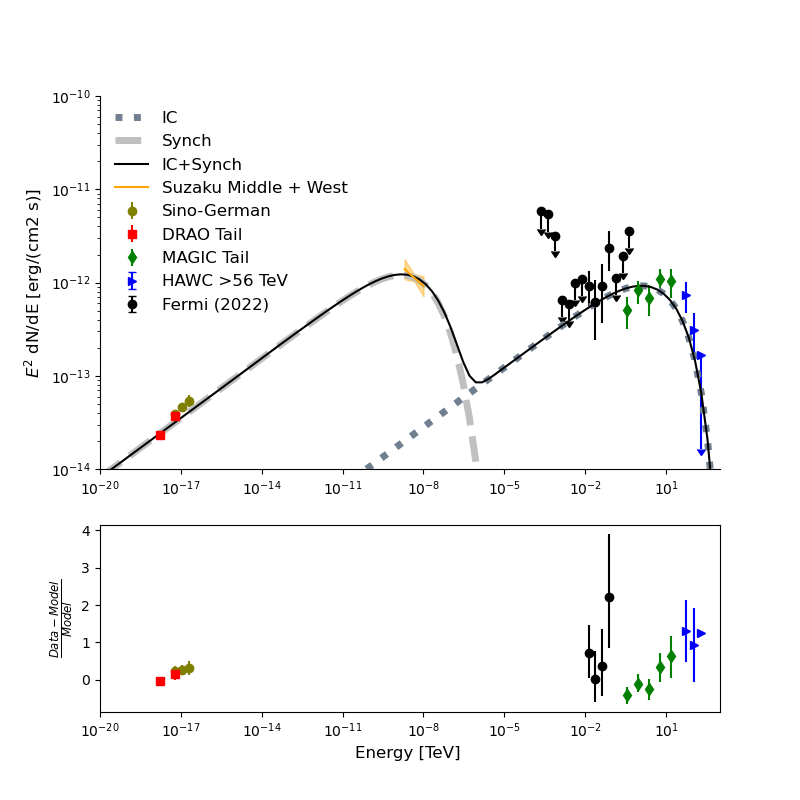}}
   \caption{Top: Multi-wavelength SED. The data shown includes the DRAO flux points in red squares \citep{p&j}, the Sino-German and Effelsberg measurements in dark green circles \citep{sino-german}, the Suzaku spectrum in yellow \citep{Ge_2021}, the \textit{Fermi}-LAT flux points in black dots \citep{fermi}, the MAGIC flux points in green diamonds \citep{magic}, and the HAWC flux points from the VHE work in blue triangles. Synchrotron is shown in dashed light grey, IC scattering is shown in dotted dark grey, and black shows the total SED from synchrotron and IC scattering combined. Bottom: Offset between the data points used and the model determined with Naima. The \textit{Fermi}-LAT upper limits are removed to better see the deviation about zero for the other data points.}
              \label{fig:naima}
    \end{figure}

The gamma ray observations are assumed to be produced through IC. For IC, the target photon fields considered are the cosmic microwave background (CMB) and near-infrared (NIR) (same as that used in \cite{tibet}). The radio and X-ray observations are assumed to be produced through nonthermal synchrotron radiation. This modeling also assumes a single population of electrons that are responsible for the radio, X-ray, and gamma ray observations. The best-fit values from the leptonic fit can be found in table~\ref{tab:naima_best-fit} and figure~\ref{fig:naima} shows the multi-wavelength model.

The leptonic modeling in this section assumed electrons are producing the UHE HAWC data which, after being accelerated in the shock fronts of the SNR ejecta or the PWN's wind, interact with the nearby photon fields and undergo IC scattering. The synchrotron and IC emission can explain the emission being seen in this region well (figure~\ref{fig:naima}). Using this multi-wavelength model, we got a total electron energy budget of 1.28$~\times~10^{45}$~erg for electrons $>$1~TeV, which is below the energy budget estimation of the SNR.

\cite{Breuhaus} suggests that electron cooling dominated by Inverse Compton (IC) losses could produce UHE emission, resulting in a PeVatron. One requirement for this system is a pulsar with a spin-down energy of $L_{SD}>10^{36}$~erg/s. While the PWN in this region fits that requirement ($L_{SD}\approx2\times10^{37}$~erg/s) and would be an ideal electron accelerator, it is offset from the location of the UHE HAWC emission and located in the head of the SNR, which we are not assuming the UHE gamma ray emission comes from. This would mean that electrons are accelerated in the shock fronts of the SN ejecta and produce gamma rays through IC. 

As mentioned in section~\ref{sec:intro}, MAGIC also saw a shift in emission towards the tail of the SNR as energy thresholds increased \citep{magic}. Even though HAWC's angular resolution is not as small as MAGIC's, we do see our UHE emission trend towards the tail of the SNR (figure~\ref{fig:vhe_maps}.b). Since we are only using energy as our differentiator for the head and tail regions of the SNR, it is also possible that electrons accelerated from the SN ejecta are producing most of the gamma ray emission, but some "leakage" from the PWN is also being accounted for in the UHE gamma ray emission. 


\section{Conclusions and outlook}
\label{sec:conclusion}

The morphological studies we performed reveal a single extended source detected over 4--145~TeV. This modeling is consistent with other gamma ray observatories. As previously stated, any discrepancies between observations are most likely attributed to the varying parts of the SNR emitting at different energies in the gamma ray regime. 

We also looked at HAWC's $>$56~TeV data to explore its correlation with nearby molecular clouds, which would hint at a hadronic gamma ray production mechanism. The molecular cloud template fits the region well and shows that protons could account for the gamma ray flux detected at these energies. We also explored the possibility of a leptonic origin for HAWC's $>$56~TeV data. While the multi-wavelength model fits all the data well, it is still hard to say what part of the region would be responsible for the UHE electrons producing gamma rays. If protons are the sole source of the UHE gamma ray emission, then SNR G106.3+2.7 is indeed a hadronic PeVatron. However, we cannot rule out the possibility of electrons producing the gamma ray emission at this time.

Currently, VERITAS, Tibet AS$\gamma$, and LHAASO also report gamma ray emission from this SNR but cannot resolve the head and tail regions of the SNR at this time. More VHE gamma ray observations disentangling the head and tail of the SNR would greatly help in describing the gamma ray emission in this region. Particularly, it would help in narrowing down which, if not both, of the regions of the SNR are capable of being a PeVatron. 


\begin{acknowledgements}
We acknowledge the support from: the US National Science Foundation (NSF); the US Department of Energy Office of High-Energy Physics; the Laboratory Directed Research and Development (LDRD) program of Los Alamos National Laboratory; Consejo Nacional de Ciencia y Tecnolog\'{i}a (CONACyT), M\'{e}xico, grants 271051, 232656, 260378, 179588, 254964, 258865, 243290, 132197, A1-S-46288, A1-S-22784, CF-2023-I-645, c\'{a}tedras 873, 1563, 341, 323, Red HAWC, M\'{e}xico; DGAPA-UNAM grants IG101323, IN111716-3, IN111419, IA102019, IN106521, IN114924, IN110521 , IN102223; VIEP-BUAP; PIFI 2012, 2013, PROFOCIE 2014, 2015; the University of Wisconsin Alumni Research Foundation; the Institute of Geophysics, Planetary Physics, and Signatures at Los Alamos National Laboratory; Polish Science Centre grant, DEC-2017/27/B/ST9/02272; Coordinaci\'{o}n de la Investigaci\'{o}n Cient\'{i}fica de la Universidad Michoacana; Royal Society - Newton Advanced Fellowship 180385; Generalitat Valenciana, grant CIDEGENT/2018/034; The Program Management Unit for Human Resources \& Institutional Development, Research and Innovation, NXPO (grant number B16F630069); Coordinaci\'{o}n General Acad\'{e}mica e Innovaci\'{o}n (CGAI-UdeG), PRODEP-SEP UDG-CA-499; Institute of Cosmic Ray Research (ICRR), University of Tokyo. H.F. acknowledges support by NASA under award number 80GSFC21M0002; National Research Foundation of Korea (RS-2023-00280210). We also acknowledge the significant contributions over many years of Stefan Westerhoff, Gaurang Yodh and Arnulfo Zepeda Dom\'inguez, all deceased members of the HAWC collaboration. Thanks to Scott Delay, Luciano D\'{i}az and Eduardo Murrieta for technical support. This research was also supported by the Nicholas Matwiyoff \& Carl Hogberg Endowed Graduate Fellowship at Michigan Technological University. 

\end{acknowledgements}

%
%

\bibliographystyle{aa.bst}
\bibliography{resources.bib}

\begin{thebibliography}{34}
\expandafter\ifx\csname natexlab\endcsname\relax\def\natexlab#1{#1}\fi

\bibitem[{Abe {et~al.}(2022)Abe, Abe, Acciari, Agudo, Aniello, Ansoldi, Antonelli, Engels, Arcaro, Artero, {et~al.}}]{magic}
Abe, H., Abe, S., Acciari, V., {et~al.} 2022, Astronomy and Astrophysics

\bibitem[{Abeysekara {et~al.}(2023)Abeysekara, Albert, Alfaro, Alvarez, Álvarez, Araya, Arteaga-Velázquez, Arunbabu, Rojas, Solares, Babu, Barber, Becerril, Belmont-Moreno, BenZvi, Blanco, Braun, Brisbois, Caballero-Mora, Martínez, Capistrán, Carramiñana, Casanova, Castillo, Chaparro-Amaro, Cotti, Cotzomi, de~León, de~la Fuente, de~León, De~Young, Hernandez, Dingus, DuVernois, Durocher, Díaz-Vélez, Ellsworth, Engel, Espinoza, Fan, Fang, Fick, Fleischhack, Flores, Fraija, García-González, Garcia-Torales, Garfias, Giacinti, Goksu, González, González-Muñoz, Goodman, Harding, Hernandez, Hernandez, Hinton, Hona, Huang, Hueyotl-Zahuantitla, Hui, Humensky, Hüntemeyer, Iriarte, Imran, Jardin-Blicq, Joshi, Kaufmann, Kieda, Kunde, Lara, Lauer, Lee, Lennarz, Vargas, Linnemann, Longinotti, Luis-Raya, Lundeen, Malone, Marandon, Marinelli, Martinez, Martínez-Castellanos, Martínez-Castro, Martínez-Huerta, Matthews, Miranda-Romagnoli, Montaruli, Morales-Soto, Moreno, Mostafá, Nayerhoda, Nellen, Newbold,
  Nisa, Noriega-Papaqui, Oceguera-Becerra, Olivera-Nieto, Omodei, Peisker, Araujo, Pérez-Pérez, Ponce, Pretz, Rho, Rosa-González, Ruiz-Velasco, Salazar, Salazar-Gallegos, Greus, Sandoval, Schneider, Schoorlemmer, Serna-Franco, Sinnis, Smith, Son, Woodle, Springer, Taboada, Tepe, Tibolla, Tollefson, Torres, Torres-Escobedo, Turner, Ureña-Mena, Ukwatta, Varela, Vargas-Magaña, Villaseñor, Wang, Watson, Werner, Westerhoff, Willox, Wisher, Wood, Yodh, Zaborov, Zepeda, \& Zhou}]{nim}
Abeysekara, A., Albert, A., Alfaro, R., {et~al.} 2023, Nuclear Instruments and Methods in Physics Research Section A: Accelerators, Spectrometers, Detectors and Associated Equipment, 1052, 168253

\bibitem[{Abeysekara {et~al.}(2017)Abeysekara, Albert, Alfaro, Alvarez, {\'{A} }lvarez, Arceo, Arteaga-Vel{\'{a}}zquez, Solares, Barber, Bautista-Elivar, Becerril, Belmont-Moreno, BenZvi, Berley, Braun, Brisbois, Caballero-Mora, Capistr{\'{a}}n, Carrami{\~{n}}ana, Casanova, Castillo, Cotti, Cotzomi, de~Le{\'{o}}n, de~la Fuente, Le{\'{o}}n, DeYoung, Dingus, DuVernois, D{\'{\i}}az-V{\'{e}}lez, Ellsworth, Fiorino, Fraija, Garc{\'{\i}}a-Gonz{\'{a}}lez, Gerhardt, Munöz, Gonz{\'{a}}lez, Goodman, Hampel-Arias, Harding, Hernandez, Hernandez-Almada, Hinton, Hui, Hüntemeyer, Iriarte, Jardin-Blicq, Joshi, Kaufmann, Kieda, Lara, Lauer, Lee, Lennarz, Vargas, Linnemann, Longinotti, Raya, Luna-Garc{\'{\i}}a, L{\'{o}}pez-Coto, Malone, Marinelli, Martinez, Martinez-Castellanos, Mart{\'{\i}}nez-Castro, Mart{\'{\i}}nez-Huerta, Matthews, Miranda-Romagnoli, Moreno, Mostaf{\'{a}}, Nellen, Newbold, Nisa, Noriega-Papaqui, Pelayo, Pretz, P{\'{e}}rez-P{\'{e}}rez, Ren, Rho, Rivi{\`{e}}re, Rosa-Gonz{\'{a}}lez, Rosenberg, Ruiz-Velasco,
  Salazar, Greus, Sandoval, Schneider, Schoorlemmer, Sinnis, Smith, Springer, Surajbali, Taboada, Tibolla, Tollefson, Torres, Ukwatta, Villase{\~{n}}or, Weisgarber, Westerhoff, Wisher, Wood, Yapici, Yodh, Younk, Zepeda, \& Zhou}]{crab2017}
Abeysekara, A.~U., Albert, A., Alfaro, R., {et~al.} 2017, The Astrophysical Journal, 843, 39

\bibitem[{Abeysekara {et~al.}(2019)Abeysekara, Albert, Alfaro, Alvarez, {\'{A} }lvarez, Camacho, Arceo, Arteaga-Vel{\'{a}}zquez, Arunbabu, Rojas, Solares, Baghmanyan, Belmont-Moreno, BenZvi, Brisbois, Caballero-Mora, Capistr{\'{a}}n, Carrami{\~{n}}ana, Casanova, Cotti, Cotzomi, de~Le{\'{o}}n, la~Fuente, de~Le{\'{o}}n, Dichiara, Dingus, DuVernois, D{\'{\i}}az-V{\'{e}}lez, Ellsworth, Engel, Espinoza, Fick, Fleischhack, Fraija, Galv{\'{a}}n-G{\'{a}}mez, Garc{\'{\i}}a-Gonz{\'{a}}lez, Garfias, Gonz{\'{a}}lez, Goodman, Harding, Hernandez, Hinton, Hona, Hueyotl-Zahuantitla, Hui, Hüntemeyer, Iriarte, Jardin-Blicq, Joshi, Kaufmann, Kieda, Lara, Lee, Vargas, Linnemann, Longinotti, Luis-Raya, Lundeen, Malone, Marinelli, Martinez, Martinez-Castellanos, Mart{\'{\i}}nez-Castro, Mart{\'{\i}}nez-Huerta, Matthews, Miranda-Romagnoli, Morales-Soto, Moreno, Mostaf{\'{a}}, Nayerhoda, Nellen, Newbold, Nisa, Noriega-Papaqui, Peisker, P{\'{e}}rez-P{\'{e}}rez, Pretz, Ren, Rho, Rivi{\`{e}}re, Rosa-Gonz{\'{a}}lez, Rosenberg,
  Ruiz-Velasco, Salazar, Greus, Sandoval, Schneider, Schoorlemmer, Arroyo, Sinnis, Smith, Springer, Surajbali, Tabachnick, Tanner, Tibolla, Tollefson, Torres, Weisgarber, Westerhoff, Wood, Yapici, Zepeda, \& Zhou}]{crab2019}
Abeysekara, A.~U., Albert, A., Alfaro, R., {et~al.} 2019, American Astronomical Society, 881, 134

\bibitem[{Abeysekara {et~al.}(2022)Abeysekara, Albert, Alfaro, Alvarez, {\'A}lvarez~Romero, Camacho, Arteaga~Velazquez, Kollamparambil, Rojas, Ayala~Solares, {et~al.}}]{hal}
Abeysekara, A.~U., Albert, A., Alfaro, R., {et~al.} 2022, in 37th International Cosmic Ray Conference.

\bibitem[{Abramowski {et~al.}(2016)Abramowski, Aharonian, Benkhali, Akhperjanian, Angüner, Backes, {et~al.}}]{hess_nature}
Abramowski, A., Aharonian, F., Benkhali, F.~A., {et~al.} 2016, Nature, 531, 476–479

\bibitem[{Acciari {et~al.}(2009)Acciari, Aliu, Arlen, Aune, Bautista, Beilicke, Benbow, Boltuch, Bradbury, Buckley, Bugaev, Butt, Byrum, Cannon, Cesarini, Chow, Ciupik, Cogan, Cui, Dickherber, Ergin, Fegan, Finley, Fortin, Fortson, Furniss, Gall, Gillanders, Gotthelf, Grube, Guenette, Gyuk, Hanna, Holder, Horan, Hui, Humensky, Kaaret, Karlsson, Kertzman, Kieda, Konopelko, Krawczynski, Krennrich, Lang, LeBohec, Maier, McCann, McCutcheon, Millis, Moriarty, Mukherjee, Ong, Otte, Pandel, Perkins, Pohl, Quinn, Ragan, Reyes, Reynolds, Roache, Rose, Schroedter, Sembroski, Smith, Steele, Swordy, Theiling, Toner, Vassiliev, Vincent, Wagner, Wakely, Ward, Weekes, Weinstein, Weisgarber, Williams, Wissel, Wood, \& Zitzer}]{veritas}
Acciari, V.~A., Aliu, E., Arlen, T., {et~al.} 2009, The Astrophysical Journal, 703, L6

\bibitem[{Ackermann {et~al.}(2017)Ackermann, Ajello, Baldini, Ballet, Barbiellini, Bastieri, Bellazzini, Bissaldi, Bloom, Bonino, {et~al.}}]{FGES}
Ackermann, M., Ajello, M., Baldini, L., {et~al.} 2017, The Astrophysical Journal, 843, 139

\bibitem[{Ade {et~al.}(2011)Ade, Aghanim, Arnaud, Ashdown, Aumont, Baccigalupi, Baker, Balbi, Banday, Barreiro, Bartlett, Battaner, Benabed, Bennett, Benoît, Bernard, Bersanelli, Bhatia, Bock, Bonaldi, Bond, Borrill, Bouchet, Bradshaw, Bremer, Bucher, Burigana, {et~al.}}]{Planck}
Ade, P. A.~R., Aghanim, N., Arnaud, M., {et~al.} 2011, Astronomy and Astrophysics, 536, A1

\bibitem[{Albert {et~al.}(2024)Albert, Alfaro, Alvarez, Andrés, Arteaga-Velázquez, Rojas, Solares, Babu, Belmont-Moreno, Caballero-Mora, Capistrán, Carramiñana, Casanova, Cotti, Cotzomi, de~León, la~Fuente, de~León, Depaoli, Lalla, Hernandez, Dingus, DuVernois, Engel, Ergin, Espinoza, Fan, Fang, Fraija, Fraija, García-González, Garfias, Goksu, González, Goodman, Groetsch, Harding, Hernández-Cadena, Herzog, Hinton, Huang, Hueyotl-Zahuantitla, Hüntemeyer, Iriarte, Kaufmann, Lara, Lee, Vargas, Linnemann, Longinotti, Luis-Raya, Malone, Martínez-Castro, Matthews, Miranda-Romagnoli, Montes, Moreno, Mostafá, Nellen, Nisa, Noriega-Papaqui, Olivera-Nieto, Omodei, Osorio, Araujo, Pérez-Pérez, Rho, Rosa-González, Ruiz-Velasco, Salazar, Salazar-Gallegos, Sandoval, Schneider, Schwefer, Serna-Franco, Smith, Son, Springer, Tibolla, Tollefson, Torres, Torres-Escobedo, Turner, Ureña-Mena, Varela, Wang, Watson, Whitaker, Willox, Wu, Yu, Yun-Cárcamo, \& Zhou}]{pass5}
Albert, A.~., Alfaro, R., Alvarez, C., {et~al.} 2024, Performance of the HAWC Observatory and TeV Gamma-Ray Measurements of the Crab Nebula with Improved Extensive Air Shower Reconstruction Algorithms

\bibitem[{Albert {et~al.}(2023)Albert, Alfaro, Alvarez, Andres, Velazquez, Rojas, Solares, Babu, Belmont-Moreno, Rojas, Yun, Carramiñana, Carreon-Gonzalez, Cotti, Cotzomi, de~León, de~la Fuente, Depaoli, de~León, Hernandez, Vélez, Dingus, Durocher, DuVernois, Engel, Hernández, Fan, Fang, Fraija, Garcia-Gonzalez, Garfias, Goksu, González, Goodman, Groetsch*, Harding, Cadena, Herzog, Hinton, Hona, Huang, Hueyotl-Zahuantitla, Hüntemeyer, Iriarte, Joshi, Kaufmann, Kieda, Lara, Lee, Lee, Vargas, Linnemann, Longinotti, Luis-Raya, Malone, Martínez-Castro, Matthews, Miranda-Romagnoli, Montes, Soto, Mostafa, Nellen, Nisa, Noriega-Papaqui, Olivera-Nieto, Omodei, Araujo, Pérez, Pratts, Rho, Rosa-Gonzalez, Ruiz-Velasco, Salazar, Salazar-Gallegos, Sandoval, Schneider, Schwefer, Serna-Franco, Smith, Yon, Springer, Tibolla, Tollefson, Torres, Escobedo, Turner, Ureña-Mena, Varela, Villaseñor, Wang, Watson, Werner, Whitaker, Willox, Wu, Zhou, \& Mora}]{4hwc}
Albert, A., Alfaro, R., Alvarez, C., {et~al.} 2023, in 38th International Cosmic Ray Conference.

\bibitem[{Albert {et~al.}(2021)Albert, Alfaro, Alvarez, Angeles~Camacho, Arteaga-Velázquez, Arunbabu, Avila~Rojas, Ayala~Solares, Baghmanyan, Belmont-Moreno, BenZvi, Brisbois, Caballero-Mora, Capistrán, Carramiñana, Casanova, Cotti, Cotzomi, Coutiño~de León, De~la Fuente, Diaz~Hernandez, Dingus, DuVernois, Durocher, Díaz-Vélez, Ellsworth, Engel, {et~al.}}]{gmc_hawc}
Albert, A., Alfaro, R., Alvarez, C., {et~al.} 2021, The Astrophysical Journal, 914, 106

\bibitem[{Albert {et~al.}(2020)Albert, Alfaro, Alvarez, Camacho, Arteaga-Vel{\'a}zquez, Arunbabu, Rojas, Solares, Baghmanyan, Belmont-Moreno, {et~al.}}]{hawc}
Albert, A., Alfaro, R., Alvarez, C., {et~al.} 2020, The Astrophysical Journal Letters, 896

\bibitem[{Amenomori {et~al.}(2021)Amenomori, Bao, Bi, Chen, Chen, Chen, Chen, Chen, Cirennima, Cui, Danzengluobu, Ding, Fang, Fang, Feng, Feng, Feng, Gao, Gou, Guo, Guo, He, He, Hibino, Hotta, Hu, Hu, Huang, Jia, Jiang, Jin, Kasahara, Katayose, Kato, Kato, Kawata, Kihara, Ko, Kozai, Labaciren, Le, Li, Li, Li, Lin, Liu, Liu, Liu, Liu, Liu, Lou, Lu, Meng, Munakata, Nakada, Nakamura, Nanjo, Nishizawa, Ohnishi, Ohura, Ozawa, Qian, Qu, Saito, Sakata, Sako, Shao, Shibata, Shiomi, Sugimoto, Takano, Takita, Tan, Tateyama, Torii, Tsuchiya, Udo, Wang, Wu, Xue, Yamamoto, Yang, Yokoe, Yuan, Zhai, Zhang, Zhang, Zhang, Zhang, Zhang, Zhang, Zhang, Zhao, Zhaxisangzhu, \& Zhou}]{tibet}
Amenomori, M., Bao, Y.~W., Bi, X.~J., {et~al.} 2021, Nature Astronomy, 5, 460–464

\bibitem[{Bell {et~al.}(2013)Bell, Schure, Reville, \& Giacinti}]{Bell_2013}
Bell, A.~R., Schure, K.~M., Reville, B., \& Giacinti, G. 2013, Monthly Notices of the Royal Astronomical Society, 431, 415–429

\bibitem[{Breuhaus {et~al.}(2021)Breuhaus, Hahn, Romoli, Reville, Giacinti, Tuffs, \& Hinton}]{Breuhaus}
Breuhaus, M., Hahn, J., Romoli, C., {et~al.} 2021, The Astrophysical Journal Letters, 908, L49

\bibitem[{Cao {et~al.}(2024)Cao, Aharonian, An, Axikegu, Bai, Bao, Bastieri, Bi, Bi, Cai, Cao, Cao, Cao, Chang, Chang, Chen, Chen, Chen, Chen, Chen, Chen, Chen, Chen, Chen, Chen, Chen, Chen, Cheng, Cheng, Cui, Cui, Cui, Cui, Dai, Dai, Dai, Danzengluobu, della Volpe, Dong, Duan, Fan, Fan, Fang, Fang, Feng, Feng, Feng, Feng, Feng, Gabici, Gao, Gao, Gao, Gao, Gao, Gao, Ge, Geng, Giacinti, Gong, Gou, Gu, Guo, Guo, Guo, Guo, Han, He, He, He, He, He, Heller, Hor, Hou, Hou, Hou, Hu, Hu, Hu, Huang, Huang, Huang, Huang, Huang, Huang, Huang, Ji, Jia, Jia, Jiang, Jiang, Jiang, Jin, Kang, Ke, Kuleshov, Kurinov, Li, Li, Li, Li, Li, Li, Li, Li, Li, Li, Li, Li, Li, Li, Li, Li, Li, Li, Li, Liang, Liang, Lin, Liu, Liu, Liu, Liu, Liu, Liu, Liu, Liu, Liu, Liu, Liu, Liu, Liu, Liu, Lu, Luo, Lv, Ma, Ma, Ma, Mao, Min, Mitthumsiri, Mu, Nan, Neronov, Ou, Pang, Pattarakijwanich, Pei, Qi, Qi, Qiao, Qin, Ruffolo, Sáiz, Semikoz, Shao, Shao, Shchegolev, Sheng, Shu, Song, Stenkin, Stepanov, Su, Sun, Sun, Sun, Tam, Tang, Tang, Tian, Wang,
  Wang, Wang, Wang, Wang, Wang, Wang, Wang, Wang, Wang, Wang, Wang, Wang, Wang, Wang, Wang, Wang, Wang, Wang, Wang, Wang, Wei, Wei, Wei, Wen, Wu, Wu, Wu, Wu, Wu, Xi, Xia, Xia, Xiang, Xiao, Xiao, Xin, Xin, Xing, Xiong, Xu, Xu, Xu, Xu, Xue, Yan, Yan, Yan, Yang, Yang, Yang, Yang, Yang, Yang, Yang, Yang, Yang, Yao, Yao, Ye, Yin, Yin, You, You, Yu, Yuan, Yue, Zeng, Zeng, Zeng, Zha, Zhang, Zhang, Zhang, Zhang, Zhang, Zhang, Zhang, Zhang, Zhang, Zhang, Zhang, Zhang, Zhang, Zhang, Zhang, Zhang, Zhang, Zhang, Zhao, Zhao, Zhao, Zhao, Zhao, Zheng, Zhou, Zhou, Zhou, Zhou, Zhou, Zhou, Zhou, Zhu, Zhu, Zhu, Zhu, \& Zuo}]{lhaaso_cat}
Cao, Z., Aharonian, F., An, Q., {et~al.} 2024, The Astrophysical Journal Supplement Series, 271, 25

\bibitem[{Cao {et~al.}(2021)Cao, Aharonian, An, Bai, Bai, Bao, Bastieri, Bi, Bi, Cai, {et~al.}}]{lhaaso}
Cao, Z., Aharonian, F., An, Q., {et~al.} 2021, Nature, 594, 33–36

\bibitem[{Cristofari(2021)}]{cristofari}
Cristofari, P. 2021, Universe, 7, 324

\bibitem[{Cristofari {et~al.}(2018)Cristofari, Gabici, Terrier, \& Humensky}]{Cristofari_2018}
Cristofari, P., Gabici, S., Terrier, R., \& Humensky, T.~B. 2018, Monthly Notices of the Royal Astronomical Society, 479, 3415

\bibitem[{Dame {et~al.}(2001)Dame, Hartmann, \& Thaddeus}]{Dame_2001}
Dame, T.~M., Hartmann, D., \& Thaddeus, P. 2001, The Astrophysical Journal, 547, 792

\bibitem[{Fang {et~al.}(2022)Fang, Kerr, Blandford, Fleischhack, \& Charles}]{fermi}
Fang, K., Kerr, M., Blandford, R., Fleischhack, H., \& Charles, E. 2022, Physical Review Letters, 129

\bibitem[{Fujita {et~al.}(2021)Fujita, Bamba, Nobukawa, \& Matsumoto}]{fujita}
Fujita, Y., Bamba, A., Nobukawa, K.~K., \& Matsumoto, H. 2021, The Astrophysical Journal Letters, 912, 133

\bibitem[{Gao {et~al.}(2011)Gao, Han, Reich, Reich, Sun, \& Xiao}]{sino-german}
Gao, X.~Y., Han, J.~L., Reich, W., {et~al.} 2011, Astronomy and Astrophysics, 529

\bibitem[{Ge {et~al.}(2021)Ge, Liu, Niu, Chen, \& Wang}]{Ge_2021}
Ge, C., Liu, R.-Y., Niu, S., Chen, Y., \& Wang, X.-Y. 2021, The Innovation, 2, 100118

\bibitem[{Halpern {et~al.}(2001)Halpern, Gotthelf, Leighly, \& Helfand}]{Halpern_2001}
Halpern, J.~P., Gotthelf, E.~V., Leighly, K.~M., \& Helfand, D.~J. 2001, The Astrophysical Journal, 547, 323–333

\bibitem[{Joncas \& Higgs(1990)}]{j&h}
Joncas, G. \& Higgs, L.~A. 1990, Astron. Astrophys. Suppl. Ser., 82, 113

\bibitem[{Kothes {et~al.}(2001)Kothes, Uyaniker, \& Pineault}]{kothes2001}
Kothes, R., Uyaniker, B., \& Pineault, S. 2001, The Astrophysical Journal Supplement Series, 560

\bibitem[{Pineault \& Joncas(2000)}]{p&j}
Pineault, S. \& Joncas, G. 2000, The Astronomical Journal, 120, 3218

\bibitem[{Reynolds(2010)}]{Reynolds_2010}
Reynolds, S.~P. 2010, Astrophysics and Space Science, 336, 257–262

\bibitem[{{Taylor} {et~al.}(2003){Taylor}, {Gibson}, {Peracaula}, {Martin}, {Landecker}, {Brunt}, {Dewdney}, {Dougherty}, {Gray}, {Higgs}, {Kerton}, {Knee}, {Kothes}, {Purton}, {Uyaniker}, {Wallace}, {Willis}, \& {Durand}}]{cgps}
{Taylor}, A.~R., {Gibson}, S.~J., {Peracaula}, M., {et~al.} 2003, \aj, 125, 3145

\bibitem[{Vannoni {et~al.}(2009)Vannoni, Gabici, \& Aharonian}]{Vannoni}
Vannoni, G., Gabici, S., \& Aharonian, F.~A. 2009, Astronomy \& Astrophysics, 497, 17–26

\bibitem[{Vianello {et~al.}(2015)Vianello, Lauer, Younk, Tibaldo, Burgess, Ayala, Harding, Hui, Omodei, \& Zhou}]{threeml}
Vianello, G., Lauer, R.~J., Younk, P., {et~al.} 2015, The Multi-Mission Maximum Likelihood framework (3ML)

\bibitem[{Zabalza(2015)}]{naima}
Zabalza, V. 2015, Proceedings of Science

\end{thebibliography}

\begin{appendix}
\section{Validating the \textit{Planck} 353 GHz dust opacity map}
\label{sec:validation}

\begin{figure*}
   \sidecaption
   \includegraphics[width=12cm]{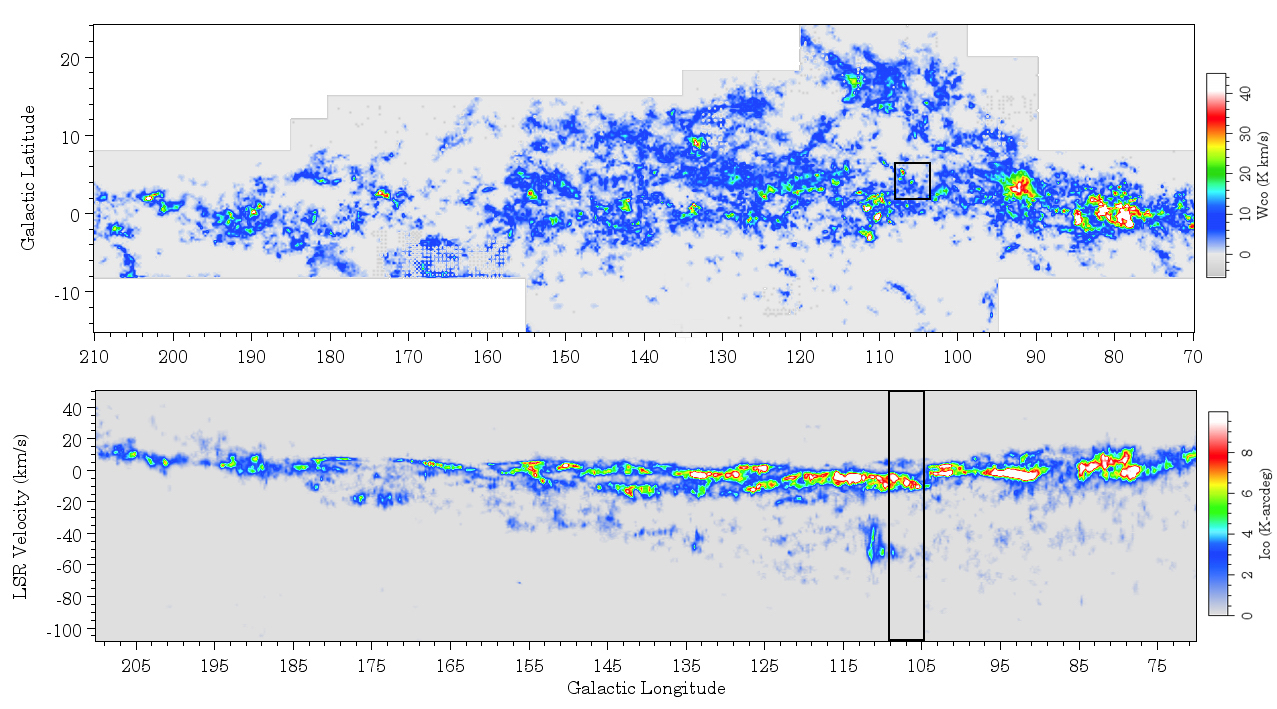}
   \caption[example] 
   { \label{fig:DH18_map} 
Dame CO survey for the Boomerang region \citep{Dame_2001}. Top: Velocity-integrated CO map. The black box indicates the Boomerang region. Bottom: Velocity range over a given longitude. The black box indicates the Boomerang longitude.}
   \end{figure*}

Two different surveys were used to assess the molecular clouds in this region: the Dame CO survey \citep{Dame_2001} and the 353 GHz \textit{Planck} dust opacity map \citep{Planck}. We used the Dame CO survey to ensure that there is a molecular cloud at the distance of the Boomerang region. Figure~\ref{fig:DH18_map} shows the velocity-integrated CO map on top, with a black box indicating the Boomerang region, and the velocity range for a given latitude on bottom\footnote{https://lweb.cfa.harvard.edu/rtdc/CO/}. From the top image, we can see that there is emission in the region and the bottom shows that there isn't any CO emission in front, or behind, the Boomerang region.

\begin{figure}
   \resizebox{\hsize}{!}{\includegraphics{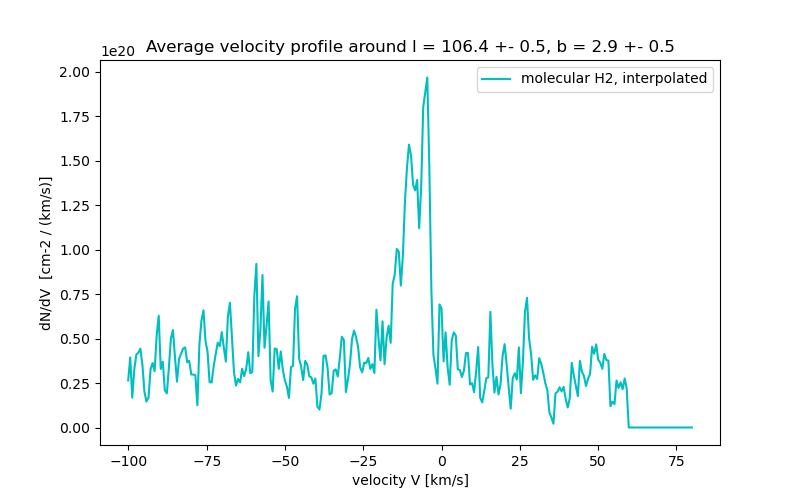}}
   \caption[example] 
   { \label{fig:velocity_distr} 
Velocity distribution of molecular hydrogen for a longitude and latitude of 106.4$^\circ$~$\pm$~0.5 and 2.9$^\circ$~$\pm$~0.5, respectively}
   \end{figure}

From here, we plotted the velocity distribution at the location of the Boomerang to see what velocity range should be integrated over in the CO map. Figure~\ref{fig:velocity_distr} shows that about $-20~\mathrm{km/s}$ to $0~\mathrm{km/s}$ is the ideal range. \cite{kothes2001} found a peak of CO emission at $-6.4~\mathrm{km/s}$, which corresponds to a distance $800~\mathrm{pc}$. We see a similar peak of emission around approximately $-6~\mathrm{km/s}$, which tells us we are seeing a similar CO emission and can use a distance of $800~\mathrm{pc}$ for the SNR and molecular cloud distances. The right shows the integrated CO map for the same velocity range. It can be seen that there is CO emission coincident with the HAWC VHE emission. 

The 353 GHz \textit{Planck} dust opacity map is newer than the Dame CO survey and provides information on HI emission, as well as CO emission because it shows gas that is optically thick in CO. Figure~\ref{fig:DH18_map} shows that there isn't any CO emission in front, or behind, the Boomerang region. This is important because the \textit{Planck} data does not take into account distance, so it is looking at everything in front, and behind, the intended target. This means that the emission we are seeing in the \textit{Planck} data is similar to the emission we are seeing in the Dame CO map at the distance of the Boomerang region and can be used for the template analysis.

\section{Fitting details for the \textit{Planck} template}
\label{sec:template}

\begin{figure*} [!htb]
   \sidecaption
   \includegraphics[width=12cm]{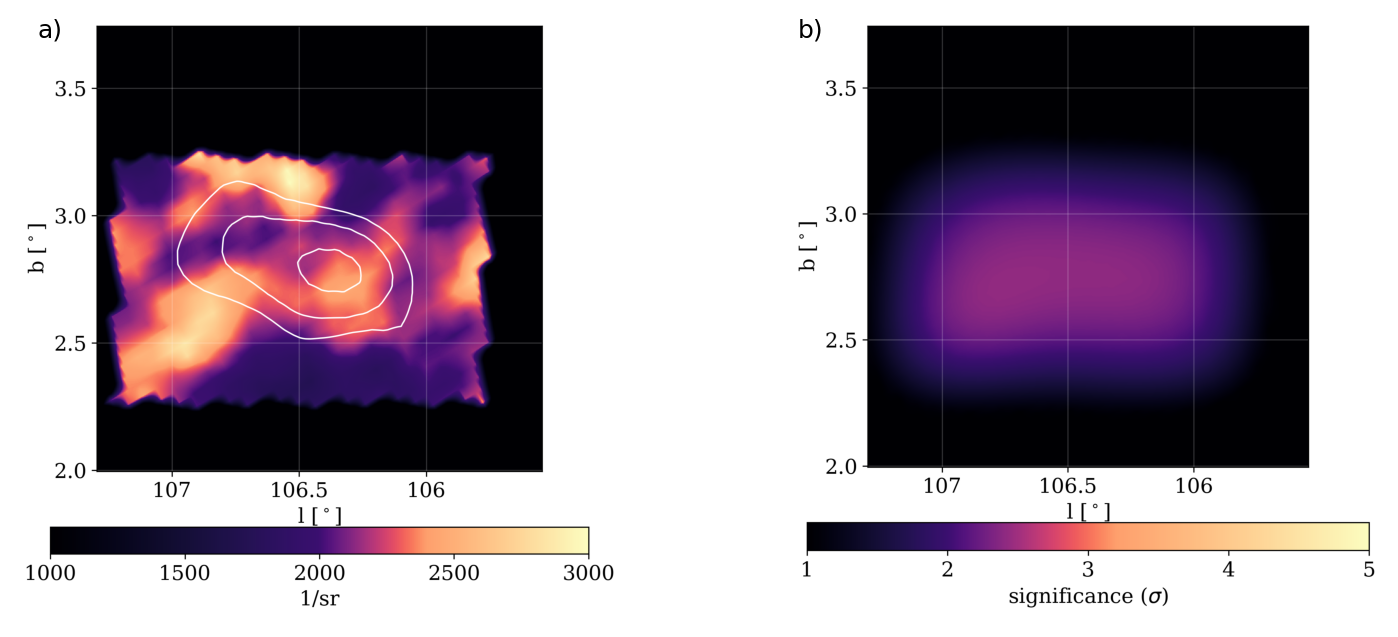}
   \caption[example] 
   { \label{fig:model_map} 
 \textbf{a)} The \textit{Planck} 353 GHz template normalized to 1/sr that is used in the VHE HAWC fit. \textbf{b)} The \textit{Planck} model template after being convolved with HAWC's PSF.}
   \end{figure*}

SNR G106.3+2.7 is a highly asymmetric source, which is not well captured by the symmetrical source assumptions that 3ML makes. We instead chose to do a template fit for the hadronic scenario to hopefully model the asymmetric source more appropriately. The template is also physically justified by the general assumption that if gamma rays are being produced through nearby molecular clouds, then the gamma ray emission might take the shape of the molecular cloud that is helping produce them. However, the \textit{Planck} data has finer angular resolutions than HAWC's point-spread function (PSF), so when the template is convolved with HAWC's PSF in the 3ML fit, we lose the fine details of the molecular clouds (figure~\ref{fig:model_map}). 

The size of the molecular cloud template is based on the size of HAWC's VHE emission, as well as the distance that would be reasonable for protons to travel for interaction. The template is 1$^{\circ}$~$\times$~1.5$^{\circ}$, which corresponds to about 14~pc~$\times$~21~pc and appropriately accounts for HAWC's emission size. Protons in this region should be able to diffuse out to about 60 pc based on \cite{hawc}, so the chosen size for the molecular cloud template is also still appropriate for proton diffusion. Since the flux fit from this template and the mass used in the CR energy density calculation is based on the size of the template we choose for the fit, the results shown here are dependent on our choices made for the template size.  

\end{appendix}
\end{document}